\documentclass[conference]{IEEEtran}
\IEEEoverridecommandlockouts

\usepackage[acronym]{glossaries}

\newglossaryentry{A_B^C}{type=main,name={\ensuremath{A_B^C}},sort=A,
description={A dummy symbol }}

\newacronym{ntnu}{NTNU}{Norwegian University of Science and Technology}
\newacronym{gui}{GUI}{Graphical User Interface}
\newacronym{vnf}{VNF}{Virtual Network Function}
\newacronym{vnfm}{VNFM}{\gls{vnf} Manager}
\newacronym{vnfd}{VNFD}{\gls{vnf} Descriptor}
\newacronym{ns}{NS}{Network Service}
\newacronym{nf}{NF}{Network Function}
\newacronym{nsd}{NSD}{\gls{ns} Descriptor}
\newacronym{nst}{NST}{Network Slice Template}
\newacronym{vim}{VIM}{Virtual Infrastructure Manager}
\newacronym{osm}{OSM}{Open Source \gls{mano}}
\newacronym{onap}{ONAP}{Open Network Automation Platform}
\newacronym{mano}{MANO}{Management and Orchestration}
\newacronym{vpn}{VPN}{Virtual Private Network}
\newacronym{mec}{MEC}{Multi-Access Edge Computing}
\newacronym{nfv}{NFV}{Network Function Virtualisation}
\newacronym{nfvi}{NFVI}{\gls{nfv} Infrastructure}
\newacronym{lcm}{LCM}{Life-Cycle Management}
\newacronym{oai}{OAI}{OpenAirInterface}
\newacronym{epc}{EPC}{Evolved Packet Core}
\newacronym{eps}{EPS}{Evolved Packet System}
\newacronym{vdu}{VDU}{Virtual Deployment Unit}
\newacronym{urllc}{URLLC}{Ultra Reliable Low Latency Communication}
\newacronym{kpi}{KPI}{Key Performance Indicator}
\newacronym{embb}{eMBB}{enhanced Mobile Broadband}
\newacronym{mmtc}{mMTC}{Massive Machine-type Communication}
\newacronym{poc}{PoC}{Proof of Concept}
\newacronym{v2x}{V2X}{Vehicle-to-Everything}
\newacronym{5gnr}{5G NR}{5G New Radio}
\newacronym{5gnsa}{NSA}{5G Non-Standalone}
\newacronym{5gsa}{SA}{5G Standalone}
\newacronym{5gc}{5GC}{5G Core}
\newacronym{5gs}{5GS}{5G System}
\newacronym{e2e}{E2E}{End-to-End}
\newacronym{vm}{VM}{Virtual Machine}
\newacronym{sdn}{SDN}{Software-Defined Networking}
\newacronym{ott}{OTT}{Over the Top}
\newacronym{mno}{MNO}{Mobile Network Operator}
\newacronym{pdu}{PDU}{Physical Deployment Unit}
\newacronym{pnf}{PNF}{Physical Network Function}
\newacronym{hnf}{PNF}{Hybrid Network Function}
\newacronym{knf}{KNF}{Kubernetes-based Network Function}
\newacronym{cnf}{CNF}{Container Network Function}
\newacronym{vpnaas}{VPNaaS}{VPN-as-a-Service}
\newacronym{lte}{LTE}{Long Term Evolution}
\newacronym{ue}{UE}{User Equipment}
\newacronym{k8s}{K8s}{Kubernetes}
\newacronym{sf}{SF}{Service Function}
\newacronym{sfc}{SFC}{\gls{sf} Chaining}
\newacronym{vca}{VCA}{VNF Configuration and Abstraction}
\newacronym{vld}{VLD}{Virtual Link Descriptor}
\newacronym{hss}{HSS}{Home Subscriber Server}
\newacronym{mme}{MME}{Mobility Management Entity}
\newacronym{spgwu}{SPGW-U}{Service Packet Gateway-User plane}
\newacronym{spgwc}{SPGW-C}{Service Packet Gateway-Control plane}
\newacronym{vcpu}{vCPU}{virtual Central Processing Unit}
\newacronym{ram}{RAM}{Random Access Memory}
\newacronym{os}{OS}{Operating System}
\newacronym{enb}{eNB}{eNodeB}
\newacronym{gnb}{gNB}{gNodeB}
\newacronym{imsi}{IMSI}{International Mobile Subscriber Identity}
\newacronym{sol}{SOL}{Solution}
\newacronym{etsi}{ETSI}{European Telecommunications Standards Institute}
\newacronym{yang}{YANG}{Yet Another Next Generation}
\newacronym{nsi}{NSI}{Network Slice Instance}
\newacronym{iaas}{IaaS}{Infrastructure-as-a-Service}
\newacronym{ran}{RAN}{Radio Access Network}
\newacronym{eutran}{E-UTRAN}{Evolved-Universal Terrestrial \gls{ran}}
\newacronym{ta}{TA}{Tracking Area}
\newacronym{sgw}{SGW}{Serving Gateway}
\newacronym{pdn}{PDN}{Packet Data Network}
\newacronym{pgw}{PGW}{\gls{pdn} Gateway}
\newacronym{ims}{IMS}{IP Multimedia System}
\newacronym{rat}{RAT}{Radio Access Technology}
\newacronym{upf}{UPF}{User Plane Function}
\newacronym{smf}{SMF}{Session Management Function}
\newacronym{amf}{AMF}{Access and Mobility Management Function}
\newacronym{cups}{CUPS}{Control and User Plane Separation}
\newacronym{sba}{SBA}{Service Based Architecture}
\newacronym{sbi}{SBI}{Service Based Interface}
\newacronym{udm}{UDM}{Unified Data Management}
\newacronym{nas}{NAS}{Non-Access Stratum}
\newacronym{as}{AS}{Access Stratum}
\newacronym{ausf}{AUSF}{Authentication Server Function}
\newacronym{nssf}{NSSF}{Network Slice Selection Function}
\newacronym{nssaaf}{NSSAAF}{Network Slice Specific Authentication and Authorization Function}
\newacronym{nrf}{NRF}{Network Repository Function}
\newacronym{nef}{NEF}{Network Exposure Function}
\newacronym{pcf}{PCF}{Policy Control Function}
\newacronym{scp}{SCP}{Service Communication Proxy}
\newacronym{qos}{QoS}{Quality-of-Service}
\newacronym{rbac}{RBAC}{Role-based Access Control}
\newacronym{cia}{CIA}{Confidentiality, Integrity and Availability}
\newacronym{vlan}{VLAN}{virtual Local Area Network}
\newacronym{api}{API}{Application Programming Interface}
\newacronym{iot}{IoT}{Internet of Things}
\newacronym{gtp}{GTP}{GPRS Tunnelling Protocol}
\newacronym{dos}{DoS}{Denial of Service}
\newacronym{ddos}{DDoS}{Distributed \gls{dos}}
\newacronym{sepp}{SEPP}{Security Edge Protection Proxy}
\newacronym{cicd}{CI/CD}{Continuous Integration/Continuous Deployment}
\newacronym{aaa}{AAA}{Authentication, Authorization and Accounting}
\newacronym{plmn}{PLMN}{Public Land Mobile Network}
\newacronym{tac}{TAC}{Tracking Area Code}
\newacronym{msin}{MSIN}{Mobile Subscription Identification Number}
\newacronym{uicc}{UICC}{Universal Integrated Circuit Card}
\newacronym{mnc}{MNC}{Mobile Network Code}
\newacronym{mcc}{MCC}{Mobile Country Code}
\newacronym{msisdn}{MSISDN}{Mobile Station International Subscriber Directory Number}
\newacronym{aka}{AKA}{Authentication and Key Agreement}
\newacronym{sapd}{SAPD}{Service Access Point Descriptor}
\newacronym{5qi}{5QI}{5G \gls{qos} Identifier}
\newacronym{kms}{KMS}{Key Management System}
\newacronym{ssh}{SSH}{Secure Shell}
\newacronym{dhcp}{DHCP}{Dynamic Host Configuration Protocol}
\newacronym{srt}{SRT}{Service Response Time}
\newacronym{mtu}{MTU}{Maximum Transmission Unit}
\newacronym{sdr}{SDR}{Software-Defined Radio}
\newacronym{sla}{SLA}{Service-Level Agreement}
\newacronym{aead}{AEAD}{Authenticated Encryption with Associated Data}

\usepackage{cite}
\usepackage{url}

\usepackage{pgfplots} 
\usepgfplotslibrary{units}
\pgfplotsset{compat=1.17}
\usetikzlibrary{patterns}
\usepackage{siunitx}
\usepackage{changepage}

\usepackage{amsmath,amssymb,amsfonts}
\usepackage{algorithmic}
\usepackage{graphicx}
\usepackage{textcomp}
\usepackage{xcolor}
\def\BibTeX{{\rm B\kern-.05em{\sc i\kern-.025em b}\kern-.08em
    T\kern-.1667em\lower.7ex\hbox{E}\kern-.125emX}}
\begin{document}


\title{Secure Service Implementation with Slice Isolation and WireGuard}
\author{\IEEEauthorblockN{Sondre Kielland, Ali Esmaeily, Katina Kralevska, and Danilo Gligoroski}
\IEEEauthorblockA{
Department of Information Security and Communication Technology\\
Norwegian University of Science and Technology (NTNU)\\
Email: \{sondrki, ali.esmaeily, katinak, danilo.gligoroski\}@ntnu.no
}
}

\maketitle

\begin{abstract}
Network slicing enables the provision of services for different verticals over a shared infrastructure. Nevertheless, security is still one of the main challenges when sharing resources. In this paper, we study how WireGuard can provide an encrypted Virtual Private Network (VPN) tunnel as a service between network functions in 5G setting. The open source management and orchestration entity deploys and orchestrates the network functions into network services and slices. We create multiple scenarios emulating a real-life cellular network deploying VPN-as-a-Service between the different network functions to secure and isolate network slices. The performance measurements demonstrate from 0.8\,Gbps to 2.5\,Gbps throughput and below 1ms delay between network functions using WireGuard. The performance evaluation results are aligned with 5G key performance indicators, making WireGuard suited to provide security in slice isolation in future generations of cellular networks.

\end{abstract}


\section{Introduction}\label{sec:introduction}
The enrollment of 5G non-standalone cellular networks is already in operation by mobile network operators. In developing Beyond 5G (B5G) networks, several planned functionalities will enable verticals to establish their services with diverse \gls{qos} requirements on shared physical infrastructure. 
Providing \gls{e2e} services over isolated network slices is a key factor to empower multiple services on a shared infrastructure. To develop agile B5G networks for supporting applications with different \gls{qos} requirements, \gls{nfv}, \gls{sdn} and \gls{mec} are the main enabling technologies~\cite{5gspillars,esmaeily2021small}.

An \gls{nfv} \gls{mano} entity connected to one or several \glspl{vim} controls and monitors the deployment of \glspl{ns} by deploying necessary infrastructure resources. For an agile network deployment, the \gls{nfv} \gls{mano} also administrates connections between \glspl{vnf}, including creation of virtual networks with the help of \gls{sdn}. Therefore, instead of manually creating and connecting the \glspl{ns} together, the \gls{nfv} \gls{mano} helps operators to deploy and control \glspl{nf} automatically. With its automatic and reusable functionality, a large number of \glspl{nf} and \glspl{ns} can be rapidly deployed on a single or multiple \glspl{vim}. 



Cloud infrastructures that can be rented or shared are necessary to utilize resources efficiently for financial and load distribution purposes. Introducing shared infrastructure raises further security challenges. Securing application data transfer over shared networks is one example of such a security challenge. A countermeasure that can be initiated against such security concerns is operating \gls{vpn} between \glspl{nf}. However, establishing \gls{vpn} tunnels introduces additional overhead. For services dependent on low latency or high throughput, the additional overhead may affect their service performance. 

\gls{nfv} \gls{mano} can provide traffic isolation for \glspl{nf} in \glspl{ns} by deploying \gls{vpn} tunneling between \glspl{nf} and interconnecting them~\cite{8104638}. In this way, the secure tunneling isolates \glspl{nsi} and the provided \glspl{ns} via the \glspl{nsi}. Nevertheless, this approach is only feasible if the \gls{vpn} does not introduce significant overhead violating \gls{qos} requirements.
The deployment of \gls{vpn} between \glspl{vnf} in an automatic mode in order to provide security isolation between slices and the effect of the introduced overhead on the performance isolation among slices in a shared environment are still open research questions. 

In this paper, we implement and analyze the performance of WireGuard for providing slice isolation in 5G environment. WireGuard~\cite{wireguard:donenfeld} is a straightforward yet immediate \gls{vpn} solution that functions via the Linux kernel and employs state-of-the-art cryptography approaches. \gls{osm} orchestrates \glspl{ns} and \glspl{nsi}, and establishes \gls{vpn} tunnels between the \glspl{vnf}. The integrated WireGuard-OSM architecture provides: 1) secure communication between the involved \glspl{vnf} of \glspl{ns} and \glspl{nsi} - slice isolation; 2) performance isolation between the slices. The performance analysis shows that the integrated WireGuard-OSM architecture meets the required \gls{kpi} values in terms of high throughput for \gls{embb} slices and low latency for \gls{urllc} slices. 
We make the code publicly available\footnote{\url{https://github.com/sondrki/TTM4905/}} to the research community.

The remainder of this paper is organized as follows. Section~\ref{sec:rw} provides a literature overview of practical approaches for secure isolation between slices. Section~\ref{sec:system-architecture} presents the system architecture. The implementation steps are explained in Section~\ref{sec:implementation}. The performance evaluation results are presented in Section~\ref{sec:performance-evaluation}. Finally, Section~\ref{sec:conclusion} concludes the paper.

\section{Related Work}
\label{sec:rw}

The isolation concept between network slices can be studied from security, performance, and dependability aspects~\cite{TheIsolationConceptinthe5GNetworkSlicing}. In addition, the \gls{cia} triad is a widely used way of looking at different security aspects. A shared infrastructure introduces security challenges in all dimensions of the \gls{cia} triad. The key feature of shared infrastructures is that an attack on or from another party sharing the infrastructure should not affect the other sharing parties. This definition of \gls{cia} is harmonic with the isolation concept in network slicing. Other parties should also be unaffected when it comes to performance and dependability, extending the availability dimension. The workload, the number of resources, and hardware or software failure of another \gls{ns} should not reduce the performance of an \gls{nf} in a separate \gls{ns} or \gls{nsi}. 

While 5G intends to fix some security issues present in the previous generations of cellular networks, it also introduces several new security threats. Some of them are raised by providing services via network slices. Paper~\cite{jisis} explores and classifies different security challenges of 5G networks. Proper isolation of logical resources is essential to avoid introducing several new risks. Eavesdropping and tampering with data, for instance, are two vectors an attacker could use to interfere with security if the application data is not properly encrypted.  
Hantouti et al. suggest that operators should deploy encrypted tunnels as a way to establish trust between \glspl{sf} to provide packet integrity and prevent bypassing of policies~\cite{sfcin5g}.



The work in~\cite{inproceedings} proposes a novel mutual authentication and key establishment protocol utilizing proxy re-encryption. The protocol grants specific authentication between components of a network slice to enable secure connection for protected key establishment among component pairs for slice security isolation. Paper~\cite{8931330} offers a secure keying scheme by adopting a multi-party computation strategy, which is appropriate for network slicing architecture in the case that third-party applications access the slices. This mechanism ensures the satisfaction of use cases or devices in which the data is collected. 

Both Haga et al. in~\cite{9289900} and Vidal et al. in~\cite{electronics10151868} focus on how a \gls{vpn} can be deployed using \gls{osm}. Reference~\cite{9289900} demonstrates how WireGuard can be added in \glspl{vnf} and compares the performance of WireGuard and OpenVPN. This proof-of-concept is carried out using two \glspl{vnf} in a single \gls{ns} with manual configuration of peer connectivity in WireGuard. For the peer setup, keys and other necessary information are obtained manually. 
Vidal et al. in~\cite{electronics10151868} uses IPsec as \gls{vpn} solution to provide link-layer connectivity for multi-site deployments. In this work, \gls{osm} deploys multiple \glspl{ns} connected through one \gls{vnf} at each \gls{nfvi}. These \glspl{vnf} handle the link layer abstraction for the other \glspl{vnf}. IPsec is used to secure the connection between the link layer providing \glspl{vnf}. Keys and connection parameters are supplied by the operator when instantiating the \gls{nsi}. 

To the best of our knowledge, none of the state-of-the-art works presents a secure service automation provisioning utilizing complex and real-life \glspl{nf}. This motivates us to integrate WireGuard tunneling with \gls{osm}, which grants secure communication between \glspl{nf} in order to establish automated and realistic network services. As a result, this system architecture guarantees security and performance isolation between \glspl{nsi}.

\section{System Architecture}\label{sec:system-architecture}

\begin{figure}[!htbp]
\centering
\includegraphics[scale=0.55]{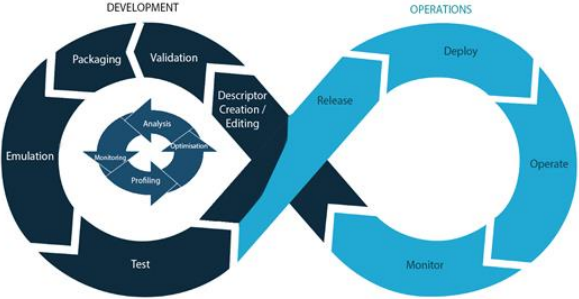}
\caption{\label{fig:servicelifecycle}Steps for service lifecycle~\cite{servicelifecycle}.}
\end{figure}

Day-0, Day-1, and Day-2 operations are terminologies used in the \gls{osm} community referring to the stages of \gls{lcm} of \glspl{nf}. The steps in Figure \ref{fig:servicelifecycle} are used to handle \gls{lcm} of \glspl{nf} via the \gls{nf} onboarding process and they link closely to Day-0 to Day-2 operations. In Figure~\ref{fig:servicelifecycle}, 
\begin{itemize} 
\item Day-0 phase focuses on necessary instantiation, including charms and descriptor creation/editing, validation, packaging, and emulation;
\item Day-1 phase concentrates on service initialization containing test, release, and deploy;
\item Day-2 phase covers runtime actions comprising operate and monitor steps.
\end{itemize}

\gls{osm} has three inbuilt supporting applications for \gls{lcm}~\cite{osminstantiation}. Cloud-init is responsible for the initial Day-0 operations like setting username and password. For Day-1 operations, Helm charts or Juju charms can be used, while Day-2 operations are also possible with Juju. The difference between Helm and Juju is that Helm is used solely for \glspl{knf}, while Juju is also usable at \gls{ns} level and for \glspl{vnf} that are not \gls{k8s} based~\cite{sol006nsd, sol006vnfd}. We have used cloud-init and Juju charms for \gls{osm} onboarding in our implementation.

Further, Juju has two operation modes: native and proxy. Native charms run operations directly inside a \gls{vnf}. On the other hand, proxy charms use a centrally placed controller, \gls{vca}, to manage the Day-1 and Day-2 actions. The \gls{vca} connects to the \glspl{vnf} through their management interface and instructs the \glspl{vnf}. The \gls{vca}-\gls{vnf} connection uses the \gls{ssh} protocol by default. In the paper, we have used proxy charms with a \gls{vca} installed co-located and integrated with \gls{osm}. Both the \gls{vca} and \gls{osm} are, therefore, able to access the \glspl{vnf} management interface to execute their actions.

To build user-defined actions, Juju uses Python scripts. The connection to the \gls{osm} instance is made through the description files of the \glspl{vnf}, \glspl{ns}, Juju config files describing metadata, and the available Day-1 and Day-2 actions. For the \gls{osm} integration of proxy charms, the \textit{charms.osm.sshproxy} library is provided by \gls{osm} to take care of, among other tasks, the basic Juju proxy peer setup. 

In addition to running actions in \glspl{vnf}, Juju can be used to create relations between Juju units for management, scaling, and handling dependencies across \glspl{vnf}. We use Juju relations to transfer WireGuard peer information between \glspl{vnf}.

\begin{figure}[!htbp]
\centerline{\includegraphics[scale=0.229]{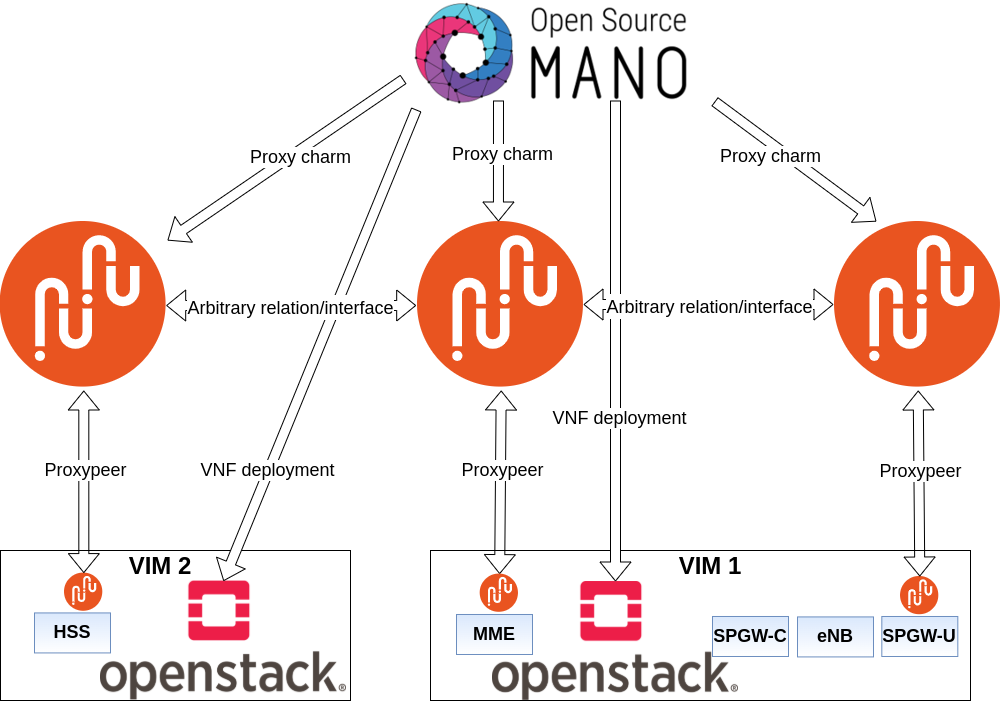}}
\caption{\label{fig:jujurelations}Interactions between elements in our Juju proxy implementation.}
\end{figure}

Figure \ref{fig:jujurelations} illustrates how we use proxy charms and relations in Juju to create a bridge for transferring information between \glspl{vnf}. The figure shows the architecture for the multi-site demonstration. Note that we used a single-\gls{vim}, moving the \gls{hss} into \textit{VIM 1}, for the performance evaluation results presented in Section~\ref{sec:implementation}. The architecture for the single-\gls{vim} setup is as illustrated in the rightmost half of the figure showing \textit{VIM 1}.




Key distribution is a task that often requires manual steps when establishing a \gls{vpn} tunnel. Manual setup can be time-consuming for dynamic environments or environments with many interfaces that need to be secured. If the tenant manager needs to do configuration, the \gls{ns} is only usable after initializing the \gls{vpn} tunnels. However, if we apply the approach presented by Vidal et al.~\cite{electronics10151868} and input the necessary information, including keys, the application can start sending data immediately after Day-1 actions have finished. Using a \gls{kms} is a similar approach. However, \gls{osm} does not provide such functionality. To use the \gls{kms} approach additional functionality outside of the \gls{osm} framework must be added.  

To perform key management, we use a non-standard approach using Juju relations with the requirement of using proxy charms for our \glspl{vnf}. By using Juju relations, we create new individual keys for every new deployment of an interface and make the application of the \gls{ns} usable directly after the Day-1 tasks finish. Furthermore, with our approach, the private keys are only stored inside the \glspl{vnf}. The public key and other necessary information for the peer setup get automatically transferred to the peer.

\section{Implementation}
\label{sec:implementation}

\begin{figure}[!htbp]
\centerline{\includegraphics[scale=0.15]{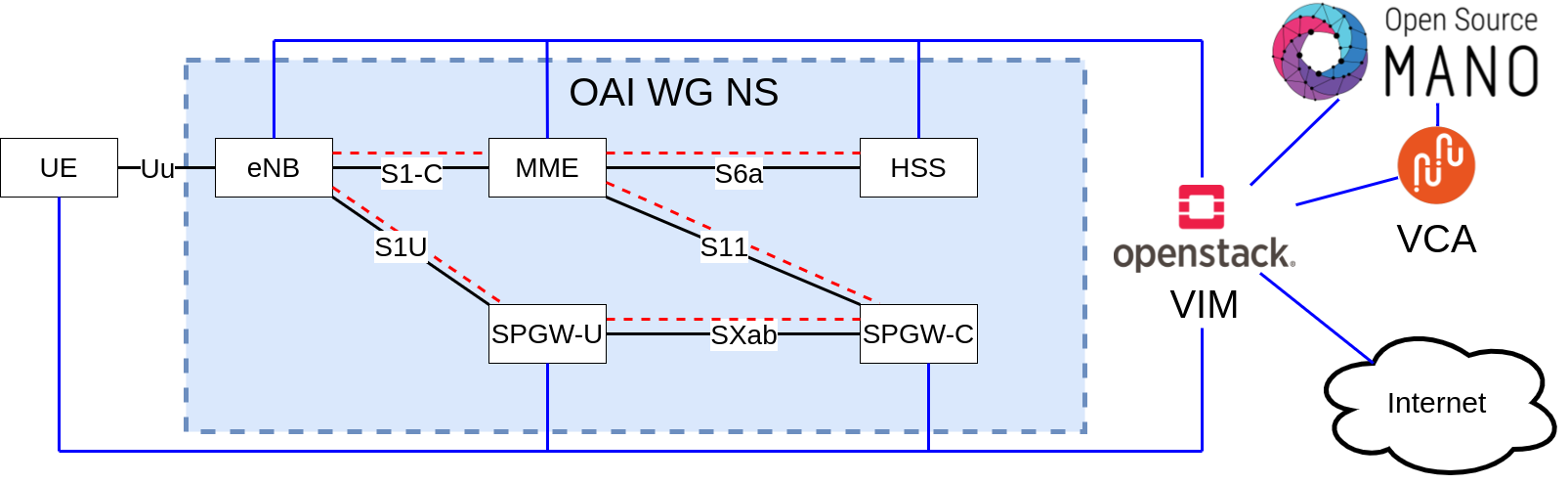}}
\caption{Architecture of our implementation.}
\label{fig:oaiepcwgarchitecture}
\end{figure}

To implement WireGuard in a realistic 5G environment we created a \gls{ns} with \gls{eps} components from \gls{oai}~\cite{oaihomepage}. We then added WireGuard connectivity on the different interfaces. Figure \ref{fig:oaiepcwgarchitecture} shows the deployed architecture. \gls{osm} is used to communicate with MicroStack \gls{vim} \cite{9165419}. The \gls{vim} hosts different \glspl{vnf}, creates virtual networks and performs routing of outgoing traffic from the \glspl{vnf}, represented by solid blue lines. WireGuard tunnel is created automatically between the \glspl{vnf} on the interfaces in the \gls{ns}, represented by the red dotted lines. In addition to the primary \gls{vim}, we utilized a second \gls{vim} in order to explore the \gls{eps} \gls{ns} deployment in multiple sites.

\subsection{Development}
We followed these steps to prepare the deployments: 1) compose a virtualized \gls{eps}, 2) set up a mechanism for automatic WireGuard peering, 3) structure \glspl{ns} into \gls{nst}, and lastly, 4) test the WireGuard connectivity in a multi-site deployment. The code for the descriptors and charms is publicly available on GitHub. In the following paragraphs, we further describe the development steps for creating the descriptors and the scripts.


\textit{1. Composing a Virtualized \gls{eps}:}
In~\cite{M2EC2020}, Dreibholz implements an \gls{epc} with \gls{hss}, \gls{mme}, and a combined \gls{sgw} and Packet Data Network Gateway (PGW) separated in two components, \gls{spgwu} and \gls{spgwc}, for user- and control-plane tasks, respectively. To extend this \gls{ns} with real-life traffic, we add a virtualized \gls{enb}. Further, we create an \gls{ue} in a \gls{vm} kept outside the \gls{ns}. The \gls{ue} is still able to connect to the \gls{enb} after instantiating the \gls{ns} with manual network setup in MicroStack. To establish the air interface, Uu, we have compiled and used \gls{oai} simulation option. When connecting the \gls{ue} to the \gls{enb}, we verify that the different \gls{eps} components function as expected and provide service to the \gls{ue}. The \gls{ue} connects to an outer network through the \gls{spgwu} via the \gls{enb}. At this first step of implementation, we still have not included WireGuard between the components.  

We chose to build the \gls{ns} by spreading the \gls{eps} components into separate \glspl{vnf}. This approach allows to split the \glspl{vnf} in the \glspl{vim}. Extending it to a multi-site environment gives us the opportunity to emulate a scenario where other components, for instance, \gls{mec}, are deployed closer to the end-users. The \glspl{vnf} distributed to remote sites are able to communicate with the core securely with the help of WireGuard.  

\textit{2. Automatic WireGuard Peering:}
Manually setting up \gls{vpn} tunnels between several interfaces can be time-consuming. Thus, we use Juju relations for automatic peering with no extra information given to the other end of the peer at the time of instantiating the \gls{ns}. The first step in the automatic peering is the establishment of relationships between \glspl{vnf} on both sides. Then the paired \glspl{vnf} retrieve information like \textit{public key}, \textit{endpoint}, and \textit{listening port} to communicate with each other. Wireguard usually employs the following cryptographic primitives: elliptic Curve25519 for key exchange, then HKDF for the key derivation, and finally, the bulk encryption work is performed by the symmetric primitive ChaCha20Poly1305 for \gls{aead}~\cite{wireguard:donenfeld}. All of these primitives have excellent performance in software supporting the objective of \gls{nfv}. Moreover, due to the lack of considerable overhead and latency, and remarkable efficiency, ChaCha20Poly1305 \gls{aead} performs significantly in terms of ping time and throughput for the \gls{urllc} and \gls{embb} slices, respectively.

To establish WireGuard connectivity on all interfaces given in Figure \ref{fig:oaiepcwgarchitecture}, we changed the IP address configuration in the components. Changing the interface addresses is necessary to route application data over the \gls{vpn} tunnel and, at the same time to ensure that applications inside the \gls{vnf} have been installed and started correctly even when waiting for the tunnel establishment. Besides, to verify that the \gls{ns} runs WireGuard, we connect the \gls{ue} and observe that it connects and gets \gls{pdn} service.

Further, in order to observe how resources affect the WireGuard performance, we have prepared a copy of the \gls{eps} \gls{ns} with WireGuard connectivity with doubled resources. 

\textit{3. NST creation:}
After having a working \gls{ns} with WireGuard connectivity between the interfaces, we include it in two \glspl{nst} to observe if and how the performance is affected by providing security with WireGuard. The two \glspl{nst} have different values of quality indicators corresponding to different \glspl{5qi}~\cite{etsi5qi}. The \gls{qos} parameters correspond to \gls{embb} and \gls{urllc} use-cases, respectively. Further, the \gls{nst} is prepared with only the management interfaces of the \glspl{vnf}. The management interfaces are attached to the external connection points in the \glspl{nst}. 

\textit{4. Multi-site deployment:}
To verify that the automatic peering setup also works in a multi-site environment, we have separated the \gls{hss} \gls{vnf} to a second \gls{vim}. When using OpenStack/MicroStack, the external floating IP address is by default not known inside a \gls{vm}. However, the \gls{vca} can retrieve the management IP address to perform its actions. To find the floating IP addresses of the \glspl{vnf}, we use the same function that Juju employs for its \textit{proxypeer} connection between a Juju unit at the \gls{vca} and the \gls{vdu} in the \gls{vnf}. After the endpoint IP address is found, the \gls{mme} and \gls{hss} connect automatically with WireGuard connectivity. A prerequisite for multi-site WireGuard connectivity is to use a port opened in the firewalls.

\subsection{Proof-of-Concept for VPN-as-a-Service}
With the automatic peering, we presented a few steps to add WireGuard as a \gls{vpnaas}. Here we summarize all steps to build the proof-of-concept. 

\begin{enumerate}
\item Append installation of WireGuard in cloud-init. 
\item Add name and parameters for Day-1 and Day-2 actions in the actions.yaml file.
\item Add relations between \glspl{vnf} in the metadata.yaml file.
\item Include the Python code to append the charm script. The name of the relationship must correspond between the name used in metadata.yaml and the listener in the \textit{\_\_init\_\_} function of the Python script.
\item Add the actions from actions.yaml into Day-1, Day-2 operations in the \gls{vnfd}. To create the WireGuard tunnel as a Day-1 operation, the relevant actions should be included in the \textit{initial-config-primitive} section in the \glspl{vnfd}. Day-2 actions are placed in the \textit{config-primitive} section.
\item While the default implementation sets up the \gls{vpn}, Day-2 actions can be used for further configuration and maintenance, for instance, if a new connection should be added towards a \gls{nf}. 

\end{enumerate}

\section{Performance Evaluation}\label{sec:performance-evaluation}


To assess the performance of WireGuard in the 5G network, we conducted measurement tests in both the control and user plane, with and without WireGuard capability. We utilized both arbitrary data and the \gls{ue} to generate realistic traffic in the network. We observe the impact of integrating secure communication with Wireguard on the performance metrics that should be aligned with the 5G \gls{kpi}~\cite{etsi5gkpi}.

While producing arbitrary data for high network load, we measure the latency and \gls{srt} in the control plane, combining multiple \gls{eps} components. In general, the following tasks are done to test the performance of \glspl{ns} and \glspl{nsi}:  

\begin{itemize} 
\item Observe \gls{srt} on the \gls{mme} when the \gls{ue} connects; 
\item Observe throughput and latency in the user plane with the \gls{ue} over S1-U interface;  
\item Measure throughput and latency between components in the \gls{eps} in the control plane over S1-C and S6a interfaces. 
\end{itemize}

\subsection{Lab Environment}
The primary \gls{vim} is a server running MicroStack with resources of 56 vCPUs, 126\,GB RAM, and 915\,GB storage. The second \gls{vim}, used for multi-site deployment, also runs MicroStack but has fewer resources with a total of 9 vCPUs, 32\,GB RAM, and 150\,GB storage. For the \gls{eps} \gls{ns} a total of 14 vCPU, 27\,GB RAM and 110\,GB storage are utilized. According to the limiting ISP, the bandwidth between the two \glspl{nfvi} is specified to be 200\,Mbps. For the \glspl{vnf} to communicate securely across the \glspl{vim}, WireGuard tunnel is established between the \glspl{nfvi}. Our measurement shows a throughput of approximately 180\,Mbps between the MicroStack instances. A nested WireGuard tunnel is used when adding WireGuard on the S6a interface for the multi-site deployment. The internal throughput of the \gls{nfvi} where the primary \gls{vim} runs is 20\,Gbps. 
Table \ref{tab:vnfinfooaienbwg} gives a summary of the resources used for the \glspl{vnf}. 

\begin{table}[!htbp]
\caption{\label{tab:vnfinfooaienbwg}VNF information of the OAI EPS NS.}
\centering
\begin{tabular}[b]{| c | c | c | c |c |}
\hline
\gls{vnf} & Operating System &
Number of & Amount of & Storage \\ 
name &  &
virtual CPUs & RAM (GB) & (GB) \\ \hline
HSS & Ubuntu18.04 & 4 & 8.0 & 20 \\ \hline
MME & Ubuntu18.04 & 2 & 4.0 & 20 \\ \hline
SPGWU & Ubuntu18.04 & 1 & 3.0 & 20 \\ \hline
SPGWC & Ubuntu18.04 & 3 & 4.0 & 30 \\ \hline
eNB & Ubuntu18.04 & 4 & 8.0 & 20 \\ \hline
UE & Ubuntu18.04 & 2 & 4.0 & 20 \\ \hline
\end{tabular} 
\end{table}

\subsection{Observations}
Before adding the \gls{vpn} tunnels, we are able to capture connection information such as the \gls{imsi}, network realms, and hostnames at the \gls{vim}. However, after we introduce WireGuard, the only information observable at the \gls{vim} is the use of the WireGuard protocol and link-layer discovery messages.  

For the control plane, we observe the \gls{srt} for the \gls{hss} application to a connecting \gls{ue}. When monitoring \gls{srt} of the \gls{hss} application including networking from the \gls{mme}, the \gls{ns} with WireGuard has the lowest average \gls{srt}. In particular, with ten successful connections for the \gls{ue}, the average \gls{srt} of the Diameter protocol drops from 6.156\,ms for the \gls{eps} without WireGuard capability to 5.377\,ms when WireGuard is added. When doubling the resources on the \gls{eps} \gls{ns} with WireGuard, \gls{srt} of 5.607\,ms is measured. Based on the other measurements, it is likely that the \gls{hss} application itself is the delaying part. With a reduced number of connections, we have not observed a negative effect on the \gls{srt} when using WireGuard. 

\begin{figure}[!htbp]
\centerline{\includegraphics[scale=0.65]{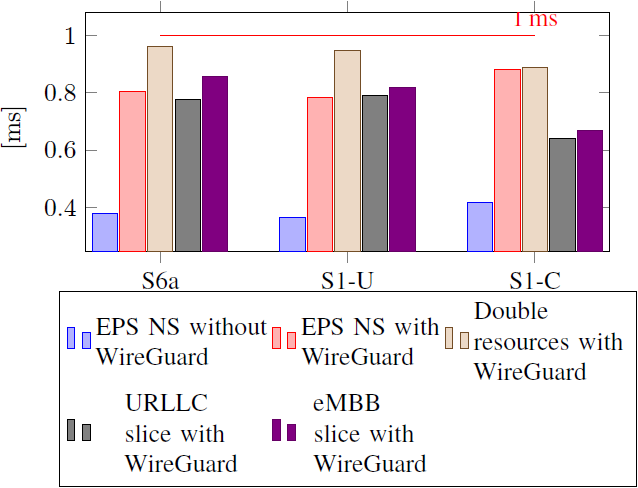}}
\caption{Latency comparison for different interfaces with and without WireGuard functionality.}
\label{fig:latcomp}
\end{figure}

A comparison of the latency measurements for different instances and interfaces is shown in Figure \ref{fig:latcomp}. The red line in the figure indicates 1\,ms, representing one of the \gls{e2e} \gls{kpi} for \gls{urllc} applications in 5G. All single-site instances achieve lower latency than the 1\,ms. However, adding WireGuard introduces a visible overhead when comparing the \gls{ns} without WireGuard to the other instances in Figure \ref{fig:latcomp}. On the other hand, we observe that the average latencies for the S1-C interface in the \gls{embb} and \gls{urllc} \glspl{nsi} (illustrated in grey and purple) are lower than the other two counterpart measurements. It is worth noting that doubling the resources does not necessarily reduce the latency, confirming that the latency depends on multiple factors such as \gls{5qi} parameters and the workload of components in the \gls{ns}.   

\begin{figure}[!htbp]
\centerline{\includegraphics[scale=0.65]{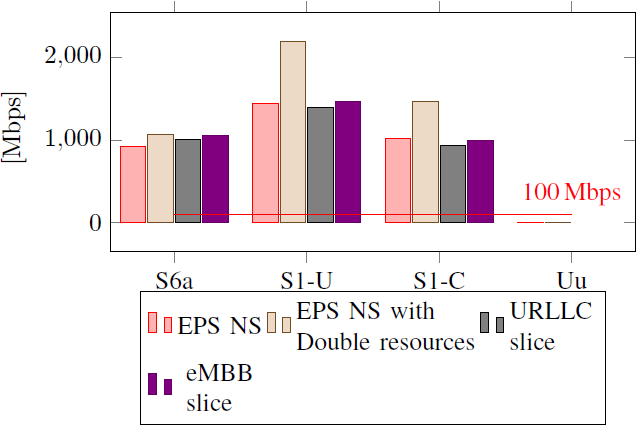}}
\caption{Throughput comparison for different interfaces with WireGuard.}
\label{fig:throughputcomp}
\end{figure}

Figure \ref{fig:throughputcomp} compares the throughput between components with WireGuard enabled on different interfaces across instances. The red line represents the 100\,Mbps downlink user data rate \gls{kpi}. We highlight three main results from observing the throughput. The first one is that, unlike the latency, the throughput changes according to the available resources. When comparing the \gls{ns} with double resources to the others, the throughput is higher for the NS with the double resources. The second observation is that the throughput over the Uu interface is significantly lower than the other measurements. The throughput over the Uu is around 1.7\,Mbps, while the average throughput for the S1-U is over 1\,Gbps making the Uu the bottleneck of the \gls{eps}. The last observation is about the maximum throughput when averaging over 10 minutes. For the \gls{ns} with double resources, we observe throughput of 2.2\,Gbps for the S1-U. For the other instances, a range from 770\,Mbps to 1.48\,Gbps is measured.  

\begin{figure}[!htbp]
\centerline{\includegraphics[scale=0.65]{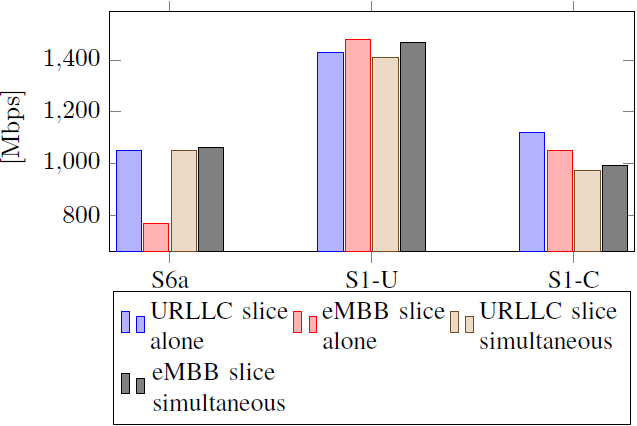}}
\caption{Throughput comparison with WireGuard for NSIs - measured separately and simultaneously.}
\label{fig:slicecomp}
\end{figure}

Figure \ref{fig:slicecomp} compares the throughput in the two \glspl{nsi}. We observe that the performance over diverse interfaces differs when running each \gls{nsi} alone and when the two \glspl{nsi} are running simultaneously. For instance, the throughput at the S6a interface reaches up to 1.1\,Gbps for the \gls{urllc} slice when it is operating alone and simultaneously with the \gls{embb} slice. However, the throughput at the S1-U interface is 1.43\,Gbps for the \gls{urllc} slice separately and it reduces a bit to 1.41\,Gbps when it is running simultaneously with the \gls{embb} slice. Regarding S1-C interface, the throughput reaches 1.12\,Gbps for the separate \gls{urllc} and it decreases to 0.97\,Gbps when the \gls{embb} slice is also working. In general, the differences between the \glspl{nsi} are minor, meaning that WireGuard is a promising solution for slice isolation of eMBB and URLLC slices.

It should be noted that we observe a total throughput of approximately 3\,Gbps, which is lower than the internal networking throughput of around 20\,Gbps when testing with a workload on the same logical interface for the two \glspl{nsi} simultaneously. As we approach the internal networking limit for the throughput, we detect more considerable differences between the \glspl{nsi} based on their \gls{qos} parameters and the allocated resources.

In the multi-site deployment, we take measurements over the S6a interface, which is the only one that differs from the other \glspl{ns} and \glspl{nsi}. As expected, the throughput is lower, and the latency is higher than in the other instances. The performance is lower even without WireGuard between the \glspl{vnf}. However, we observe that WireGuard adds overhead in this scenario as well. In the multi-site \gls{ns}, the average latency over 1000 ICMP packets increases from 18.355\,ms to 19.769\,ms when using WireGuard. For the average throughput, we observe a reduction from 179\,Mbps to 156\,Mbps, which is expected based on the given 200\,Mbps bandwidth.

\section{Conclusions}\label{sec:conclusion}
By using Juju relations and providing a proof of concept for using WireGuard as \gls{vpnaas}, we showed that WireGuard can be implemented with automatic peer setup after instantiating. The performance measurements demonstrate that WireGuard is suitable for applications with requirements corresponding to several of the 5G \gls{kpi} values. We show that WireGuard can be used as \gls{vpnaas} in the context of 5G networks and beyond in order to provide secure communication and slice isolation. 

Replacing the arbitrary Juju relations with a \gls{kms}, using a 5G Core network instead of \gls{epc} components, adding multiple \glspl{ue}, and evaluating scenarios in which fulfilling service requirements (especially throughput) are beyond WireGuard capability are potential directions for future investigation.

\bibliographystyle{IEEEtran}
\bibliography{main}





\end{document}